\makeatletter\def\graphicscache@inhibit{true}\makeatother
\PassOptionsToPackage{hashshortnames}{graphicscache}

\documentclass[letterpaper, 10 pt, conference]{ieeeconf}  %

\IEEEoverridecommandlockouts                              %

\usepackage[utf8]{inputenc}

\usepackage[hidelinks]{hyperref}

\usepackage[style=ieee,hyperref,natbib=true,backend=biber,firstinits,doi=false,%
     mincitenames=1,maxcitenames=2,maxbibnames=6,sorting=none,terseinits=false,hyperref=true]{biblatex}
\bibliography{references.bib}
\renewbibmacro*{bbx:savehash}{}%
\defbibheading{bibliography}[\bibname]{\section*{References}}

\AtEveryBibitem{\clearfield{pages}\clearfield{publisher}}

\usepackage{url}
\usepackage{graphicx}
\usepackage{grffile}
\usepackage[usenames,dvipsnames]{xcolor}
\usepackage[draft,nomargin,inline]{fixme}
\fxsetup{theme=color}

\usepackage{amsmath}
\usepackage{amssymb}
\usepackage{textcomp}
\usepackage{siunitx}

\usepackage{booktabs}
\usepackage{threeparttable}
\usepackage{multirow}

\usepackage[capitalize]{cleveref}

\usepackage{tikz}
\usetikzlibrary{arrows}
\usetikzlibrary{positioning,calc}
\usetikzlibrary{decorations.pathreplacing}
\usetikzlibrary{decorations.markings}
\usetikzlibrary{fit}
\usetikzlibrary{shapes.callouts}
\usetikzlibrary{shapes.geometric}
\usetikzlibrary{matrix}
\usetikzlibrary{arrows}
\usetikzlibrary{decorations.pathmorphing}
\usetikzlibrary{calligraphy}
\usetikzlibrary{shapes.arrows}
\usetikzlibrary{spy}

\pgfdeclarelayer{background}
\pgfdeclarelayer{foreground}
\pgfsetlayers{background,main,foreground}

\makeatletter

\tikzdeclarecoordinatesystem{rel}{%
	\tikzset{cs/.cd,x=0pt,y=0pt,#1}%
	\pgfpointlineattime{(\tikz@cs@x)}%
	{\pgfpointanchor{\tikz@pp@name{\tikz@cs@node}}{south west}}%
	{\pgfpointanchor{\tikz@pp@name{\tikz@cs@node}}{south east}}%
	\edef\tikz@cs@x{\the\pgf@x}%
	\pgfpointlineattime{(\tikz@cs@y)}%
	{\pgfpointanchor{\tikz@pp@name{\tikz@cs@node}}{south west}}%
	{\pgfpointanchor{\tikz@pp@name{\tikz@cs@node}}{north west}}%
	\pgfpoint{\tikz@cs@x}{\pgf@y}%
}

\newcommand\currentcoordinate{\the\tikz@lastxsaved,\the\tikz@lastysaved}

\makeatother

\usepackage{pgfplotstable}
\usepackage{pgfplots}
\pgfplotsset{compat=1.15}
\usepgfplotslibrary{groupplots}
\usepgfplotslibrary{units}
\usepgfplotslibrary{statistics}
\usepgfplotslibrary{colorbrewer}
\usepgfplotslibrary{polar}

\tikzset{
  /pgfplots/colormap={greenred}{
      color=(RdYlGn-O);
      color=(RdYlGn-N);
      color=(RdYlGn-L);
      color=(RdYlGn-J);
      color=(RdYlGn-I);
      color=(RdYlGn-H);
      color=(RdYlGn-G);
      color=(RdYlGn-F);
      color=(RdYlGn-D);
      color=(RdYlGn-B);
      color=(RdYlGn-A);
  },
}

\usepgfplotslibrary{external}

\tikzset{
  densitylabel/.style={font=\scriptsize, fill=white, inner sep=0pt}
}
\pgfplotsset{
  density/.style={
    mesh/ordering=y varies, colormap/OrRd, set layers=axis on top,
    xtick=\empty,
    xtick style={draw=none},
    ytick style={draw=none},
    xticklabel=\empty,
    yticklabel=\empty,
    y axis line style={opacity=0},
    ymax=45,
    ytick={10,20,30,40},
    scale only axis,
    height=2.45cm
  }
}

\usepackage{blindtext}

\usepackage[export]{adjustbox}

\usepackage{placeins}
\usepackage{dblfloatfix}

\usepackage{blindtext,letltxmacro,xcolor,xparse}
\LetLtxMacro{\blindtextblindtext}{\blindtext}
\LetLtxMacro{\blindtextBlindtext}{\Blindtext}

\RenewDocumentCommand{\blindtext}{O{\value{blindtext}}}{%
  \begingroup\color{gray}\blindtextblindtext[#1]\endgroup
}

\IfFileExists{graphicscache.sty}{%
    \usepackage{graphicscache}%
    \newcommand{\includenocompress}[2][]{\includegraphics[compress=false,##1]{##2}}%
}{%
    \newcommand{\includenocompress}[2][]{\includegraphics[##1]{##2}}%
}

\makeatletter
\newcommand\coord{\the\tikz@lastxsaved,\the\tikz@lastysaved}
\makeatother

\tikzset{
  font=\sffamily\footnotesize,
  m/.style={draw, rounded corners, fill=yellow!20, align=center}
}

\makeatletter
\tikzset{
  border on top/.style={
    append after command={
      \bgroup
      node [rectangle,fit=(\tikzlastnode),inner sep=-\pgflinewidth,draw,border on top prevent] {}
      \egroup
    },
  },
  border on top prevent/.code={%
      \let\tikz@after@path\pgfutil@empty%
  },
}
\makeatother

\usepackage{subfiles}

\title{\LARGE \bf
Foveated Compression for Immersive Telepresence Visualization
}

\author{Max Schwarz$^{*}$ and Sven Behnke%
\thanks{$^{*}$Both authors are with the Autonomous Intelligent Systems group of University of Bonn, Germany; the Lamarr Institute for Machine Learning and Artificial Intelligence; and the Center for Robotics at University of Bonn. Contact: {\tt schwarz@ais.uni-bonn.de}}%
}

\begin{document}

\maketitle

\begin{abstract}

Immersive televisualization is important both for telepresence and teleoperation, but
resolution and fidelity are often limited by communication bandwidth constraints.
We propose a lightweight method for foveated compression of immersive televisualization
video streams that can be easily integrated with common video codecs, reducing
the required bandwidth if eye tracking data is available.
Specifically, we show how to spatially adjust the Quantization Parameter of modern
block-based video codecs in a adaptive way based on eye tracking information.
The foveal region is transmitted with high fidelity while quality is reduced in the peripheral region, saving bandwidth.
We integrate our method with the NimbRo avatar system, which won the
ANA Avatar XPRIZE competition.
Our experiments show that bandwidth can be reduced to a third
without sacrificing immersion. We analyze transmission fidelity with qualitative examples
and report quantitative results.

\end{abstract}

\section{Introduction}

Telerobotics is an important field in robotics which receives
increasing attention in recent years~\cite{StotkoKSLBKW19, ZhouZD20, KlamtSLBBCDDGKK20,  WalkerPC0S23, DarvishPRCPYIP23, BehnkeAL23, WuSYLA24GELLO}, fueled by the availability of high-quality
sensors and displays, high-bandwidth network connections, and advances in haptics
and audiovisual telepresence.
Increasingly, teleoperation is used to teach robots complex capabilities, producing
 teacher signals for learning~\citep{SiWY21,zhao2023learning,MobileALOHA2024}.
\begin{tikzpicture}[remember picture,overlay]
  \node[anchor=north,align=center,font=\sffamily\small,yshift=-0.4cm] at (current page.north) {%
  \textbf{Accepted final version.} IEEE Conference Telepresence (TELEPRESENCE), Leiden, Netherlands, September 2025};
\end{tikzpicture}%

Telepresence systems must capture the remote scene and display it to the operator in an immersive way.
To this end, the most important human sense to be transmitted is vision. Since the human visual 
system is capable of resolving very fine details, high resolution image data needs to
be captured on the robot side and displayed to the remote operator.
The communication network between robot and operator is often severely limited
in bandwidth, though. For example, a search \& rescue robot inside a building might have to
rely on low-bandwidth wall-penetrating wireless technology.

However, the human eye has variable acuity, with most resolution dedicated to
the central region, the fovea~\citep{wang2023foveated}.
In the periphery, resolution is drastically reduced---meaning
that any high-resolution image data displayed there serves no purpose.
This insight is not new, especially in VR contexts, where foveated rendering~\citep{wang2023foveated}
and foveated video compression and streaming~\citep{kamarainen2023foveated,pohl2017next,kaplanyan2019deepfovea,bezugly2021lightweight}
are common techniques to reduce bandwidth and/or computational load.

In this work, we propose to use foveated compression for televisualization in the context of robotic telepresence systems
to reduce bandwidth demands without adverse effects on immersion.
To this end, we use eye tracking information to predict the operator's gaze direction,
project it into the camera image using kinematic information,
compute a Quantization Parameter (QP) delta map and feed it into the HEVC video codec
to increase lossy compression in the periphery.

We integrate our method in the NimbRo avatar system~\citep{lenz2025nimbro}, the system
that won the ANA Avatar XPRIZE Challenge~\citep{hauser2024analysis}. It features
a movable 6-DoF head with a high-resolution stereo camera system.

We  evaluate the integrated system with qualitative and quantitative experiments.
Our results show that bandwidth can be reduced significantly
without sacrificing immersion. We analyze transmission fidelity 
and report performance metrics.

\begin{figure}
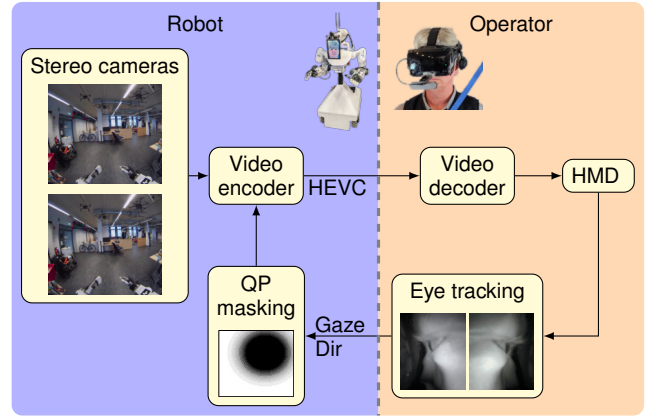

 \centering
 \begin{tikzpicture}
  \matrix[matrix of nodes, nodes={m,align=center,anchor=center}, column sep=0.25cm, row sep=-0.5cm] (matrix) {
    \node (cams) {Stereo cameras\\[1ex]\includegraphics[width=1.5cm]{images/fov/feed.png}\\[1ex]\includegraphics[width=1.5cm]{images/fov/feed.png}}; & \node (enc) {Video\\encoder}; &[0.9cm] \node (dec) {Video\\decoder}; & |(hmd)| HMD \\
                            & \node (qp) {QP\\masking\\[1ex]\includegraphics[frame,width=1cm]{images/fov/focus.png}};     & \node (eye) {Eye tracking\\[1ex]\includegraphics[height=1cm,clip,trim=0 100 0 100]{images/vr_cams/left.png}\,\includegraphics[height=1cm,clip,trim=0 100 0 100]{images/vr_cams/right.png}}; \\
  };

  \draw[-latex] (cams) -- (enc);
  \draw[-latex] (enc) -- node[pos=0.01,below right,inner sep=1pt] {HEVC} (dec);
  \draw[-latex] (dec) -- (hmd);
  \draw[-latex] (qp) -- (enc);
  \draw[-latex] (eye) -- node[pos=0.99,right,align=left]{Gaze\\Dir} (qp);
  \draw[-latex] (hmd) |- (eye);

  \coordinate (mid) at ($(qp.east)!0.85!(eye.west)$);

  \coordinate (label) at ($(matrix.north-|mid)+(0,0.1)$);
  \coordinate (rlabel) at ($(label)!0.5!(matrix.west)$);
  \node[anchor=base, inner sep=5pt] at (rlabel|-label) {Robot};

  \coordinate (olabel) at ($(label)!0.5!(matrix.east)$);
  \node[anchor=base, inner sep=5pt] (oplabel) at (olabel|-label) {Operator};

  \draw[very thick,dashed,draw=black!50] (oplabel.north-|mid) -- (matrix.south-|mid);

  \node[anchor=north east] at (oplabel.north-|mid) {\includegraphics[height=1.6cm]{images/anna.png}};
  \node[anchor=north west] at (oplabel.north-|mid) {\includegraphics[height=1.3cm]{images/jerry_head.png}};

  \begin{pgfonlayer}{background}
   \fill[blue!30] (oplabel.north-|mid) [rounded corners] -| (matrix.south west) [sharp corners] -- (matrix.south-|mid) -- cycle;
   \fill[orange!30] (oplabel.north-|mid) [rounded corners] -| (matrix.south east) [sharp corners] -- (matrix.south-|mid) -- cycle;
  \end{pgfonlayer}
 \end{tikzpicture}
 \caption{Overview of our method.
 We provide immersive visualization of the robot's environment to the operator by streaming stereo images to the operator's HMD.
 For transmission over a potentially bandwidth-limited network, the video stream is encoded using HEVC.
 We propose to utilize eye tracking information to modulate HEVC's Quantization Parameter (QP) spatially---decreasing quality (and thus bandwidth usage) in areas that are not in the focus of the operator.
 }
 \label{fig:teaser}
\end{figure}

In short, our contributions include:
\begin{enumerate}
  \item A lightweight technique for foveated compression that can be easily integrated with common video codecs,
  \item integration of the method with the \mbox{NimbRo} avatar system, and
  \item qualitative and quantitative experiments demonstrating its effectiveness.
\end{enumerate}

\section{Related Work}
Related works comprise saliency-based video compression, foveated rendering, and foveated streaming.

\paragraph{Saliency-Based Video Compression}
A larger corpus of related work focuses on predicting human visual attention
in order to focus compression quality on attended regions~\citep{itti2004automatic,lyudvichenko2017semiautomatic}.
Usually, no eye tracking is involved and the focus point is generated purely
from predicted saliency in the image.
\Citet{lyudvichenko2017semiautomatic} describe a hybrid method that
combines eye-tracking data captured from a human observer with a model of spatiotemporal
attention to predict multiple focus points.
In contrast to our method, the above works focus on offline video compression
and are thus not directly suitable for real-time usage.

\paragraph{Foveated Rendering}
Foveated Rendering aims to reduce computational load for rendering VR scenes by
reducing rendering quality in the peripheral areas. \Citet{wang2023foveated} provide
an overview of the field.
Similar to our method, foveated rendering utilizes eye tracking hardware built into the HMD
to estimate gaze direction.
While foveated rendering can be employed for VR telepresence, it can only
reduce computational load on the operator side.

\paragraph{Foveated Streaming}
In order to reduce network bandwidth, the quality reduction needs to happen earlier
in the pipeline on the producer side (as in our work, see \cref{fig:teaser}).
\Citet{kamarainen2023foveated} follow this idea, but for remote rendering, where
VR scene generation is moved to the cloud instead of the client.
\Citet{pohl2017next} also describe a foveated compression scheme for (2D) video streaming,
where frames are divided into nine rectangular regions according to eye tracking information.
The central region is transmitted with full resolution, while resolution is reduced for the periphery.
In contrast, our method allows much more fine-grained spatial control of compression quality.
\Citet{kaplanyan2019deepfovea} propose DeepFovea, a method for reconstructing
video from a foveated encoding that retains only sparse color information
in the periphery. In contrast to our method, which works with common video codecs,
it is not immediately clear how to compress and transmit video efficiently with this
representation.
\Citet{bezugly2021lightweight} propose a lightweight pre- and postprocessing stage
that offers foveated compression around a video encoding. It works by warping
the input image in such a way that the foveal region is retained in its original form,
but the periphery is compressed into less space. The warped image is encoded and
decoded as usual by the video codec and then de-warped.
This approach controls bandwidth by varying resolution instead of compression levels.
However, it suffers from warping artifacts such as aliasing effects.
\Citet{lungaro2018gaze} implement foveated compression for video transmission
by overlaying a low-resolution background stream with high-resolution foreground
tiles selected by the user's gaze. In contrast, our method offers higher spatial
resolution of quality control.
\citet{WuLC0Z24Theia} present gaze-driven perception-aware volumetric content
delivery for mixed reality headsets employing a log-polar transformation around the fixation point for 2D content augmentation. 
\citet{LiDBBV21} introduce a log-rectilinear transformation to enable foveated streaming of 360° videos with off-the-shelf video codecs for VR headsets with eye-tracking.
\citet{illahi2021foveated} survey foveated rendering, encoding, and warping techniques for foveated streaming.
In earlier work, \citet{illahi2020cloud} propose a very similar QP modulation scheme as the one presented in this
paper, but applied to video streaming for remote gaming.
\Citet{Tefera2024} use foveated compression for \mbox{RGB-D} point clouds in a third-person robot telemanipulation setting. However, point clouds are difficult to display
in dense fashion, especially when subsampled. Furthermore, \mbox{RGB-D} sensors
often fail on reflective or transparent materials.

To our knowledge, there is no prior work on foveated video compression
for immersive telepresence in robots.

\section{Method}

\subsection{NimbRo Avatar Telepresence System}

We base our work on our NimbRo Avatar system~\citep{lenz2025nimbro}, which consists
of an anthropomorphic robotic avatar and an accompanying operator station and strives for full
audiovisual and haptic immersion. It was designed for the ANA Avatar XPRIZE competition~\citep{BehnkeAL23,hauser2024analysis},
where our team received the highest score in the semifinals and won the grand prize in the finals.

\subsection{High-resolution Wide-angle Stereo Cameras}

The avatar robot is equipped with a 6-DoF-movable head with two Basler a2A3840-45ucBAS
cameras in stereo configuration. Fisheye lenses are mounted on the cameras to provide
a field of view of more than $180^\circ$. Images are captured at 47\,Hz with a resolution
of 2472$\times$2178 pixels and processed in real-time in the onboard computer.

The images are captured and transferred in a native 12-bit Bayer format to the Nvidia RTX 3070
GPU, where they are processed using CUDA. After unpacking, automatic white balancing,
and debayering, the RGB images are passed through a bilateral filter with $\sigma_d=1.1$ and $\sigma_r=0.01$
to reduce remaining debayering artifacts.
After exposure and tone mapping (ACES Film), we perform color correction in the Oklab color space~\citep{ottosson2020oklab}
and finally convert to the sRGB color space through gamma correction.
The operations are implemented as CUDA kernels and fused as far as possible to guarantee minimum latency.

\subsection{Video Codec \& Network Transmission}

The video frames are encoded using the Nvidia NVENC HEVC codec directly on the GPU.
We utilize NVENC's \texttt{ULTRA\_LOW\_LATENCY} preset to achieve minimum latency and use
intra refresh rather than IDR frames to guarantee constant bandwidth.
The compressed frames are then transferred to the CPU and transmitted using
the \texttt{nimbro\_network} library which provides robust network transport over unreliable
links for the ROS1 ecosystem\footnote{\url{https://github.com/AIS-Bonn/nimbro_network}}.
For details on the NimbRo avatar system's WiFi transmission, we refer to \citet{lenz2025nimbro}.

On the operator side, the compressed video frames are received and uploaded to the GPU
of the operator PC, an Nvidia RTX A6000. The frames are decoded using the built-in NVDEC hardware
decoder and directly passed to the VR rendering component.

\subsection{HMD \& Eye Tracking}

\begin{figure}
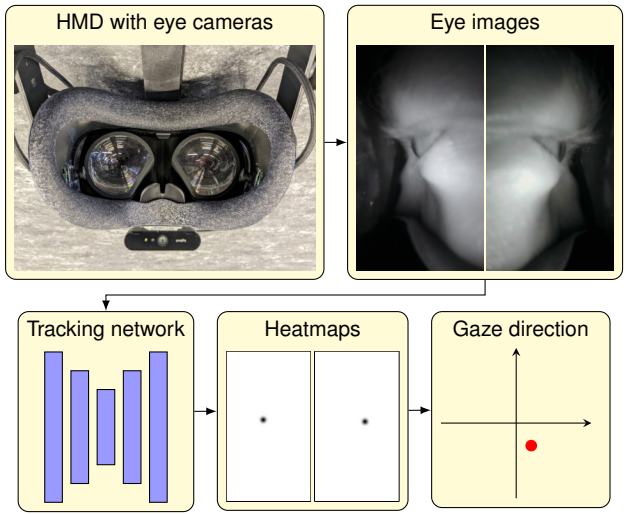

 \centering
 \newsavebox\hourglass
 \sbox\hourglass{%
   \begin{tikzpicture}[
     layer/.style={minimum height=1cm,minimum width=0.1cm,fill=blue!40,draw}
   ]
    \node[layer, minimum height=2cm] (d1) {};
    \node[layer, minimum height=1.5cm, right=0.1cm of d1] (d2) {};
    \node[layer, minimum height=1cm, right=0.1cm of d2] (d3) {};
    \node[layer, minimum height=1.5cm, right=0.1cm of d3] (u2) {};
    \node[layer, minimum height=2cm, right=0.1cm of u2] (u3) {};
   \end{tikzpicture}%
 }
 \newsavebox\gazeplot
 \sbox\gazeplot{%
   \begin{tikzpicture}\begin{axis}[
        ymin=-1,ymax=1,xmin=-0.9,xmax=0.9, scale only axis, width=2cm, height=2cm,
        axis equal,xtick=\empty,ytick=\empty,
        axis lines=center
   ]
   \addplot[red,only marks] table [x index=0, y expr=-\thisrowno{1},row sep=\\] {
        0.2 0.3 \\
   };
  \end{axis}\end{tikzpicture}
 }
 \begin{tikzpicture}[
   layer/.style={minimum height=1cm,minimum width=0.1cm,fill=blue!40,draw},
   every outer matrix/.style={inner sep=0cm}
  ]
  \matrix[column sep=0.3cm, row sep=0.3cm] (m1) {
    \node[m] (hmd) {HMD with eye cameras\\[1ex]\includegraphics[height=3cm,angle=180]{images/vr.jpg}}; &
    \node[m] (eye) {Eye images\\[1ex]
      \includegraphics[height=3cm]{images/vr_cams/left.png}\,%
      \includegraphics[height=3cm]{images/vr_cams/right.png}
    }; \\
  };
  \matrix[below=.4cm of m1, column sep=0.3cm] {
    \node[m] (net) {Tracking network\\[1ex]\usebox\hourglass}; &
    \node[m] (maps) {Heatmaps\\[1ex]
      \includegraphics[frame,height=2cm]{images/vr_cams/heatmap_left.png}\,%
      \includegraphics[frame,height=2cm]{images/vr_cams/heatmap_right.png}
    }; &
    \node[m] (out) {Gaze direction\\[1ex]\usebox\gazeplot}; \\
  };
  \draw[-latex]
    (hmd) edge (eye)
    (eye.south) -- ++(0,-0.2) -| (net.north)
    (net) edge (maps)
    (maps) edge (out);
 \end{tikzpicture}
 \caption{Eye tracking pipeline. We mount additional eye cameras into an
 off-the-shelf HMD (Valve Index).
 The captured eye images are processed by a lightweight hourglass network
 to heatmaps describing an eye keypoint.
 Finally, the detected keypoints are transformed into the eye coordinate system and averaged to determine the gaze direction.
 }
 \label{fig:eye}
\end{figure}

In our work, we use a Valve Index HMD, offering 2$\times$1440$\times$1600
resolution at up to 144\,Hz with $108^\circ$ horizontal field of view.
In earlier work~\citep{rochow2022vr}, we modified the HMD to add eye cameras
and a mouth camera to capture facial dynamics of the person wearing it.

For clarity, we briefly explain the eye tracking component here (see \cref{fig:eye}). We mounted two
miniature cameras inside the HMD that look from the sides inwards. Illumination
is provided by IR LEDs without distracting the user.
The system is initially calibrated by having the user follow a moving red dot
with their eyes. We then train a lightweight hourglass network with only two
downsampling and two upsampling layers that processes each eye camera image
and outputs a heatmap describing the eye keypoint location.
This keypoint is supervised only through a learned transformation
from the camera coordinate system into the eye coordinate system, where
calibration data is available.
Calibration and network training can be done in roughly one minute.

During operation, predicted gaze directions from left and right eye are averaged
to produce a single focus ray.

Our method does not depend on details of this eye tracking method.
Therefore, off-the-shelf eye tracking systems can also be used.

\subsection{6-DoF Camera Movement \& Spherical Rendering}

\begin{figure}
 \centering
  \begin{tikzpicture}[font=\footnotesize]
  \node[anchor=south,label={below:(a) 6-DoF head with stereo cameras}] at (0,-1) {\includegraphics[frame,height=3.5cm,clip,trim=0 10 10 20]{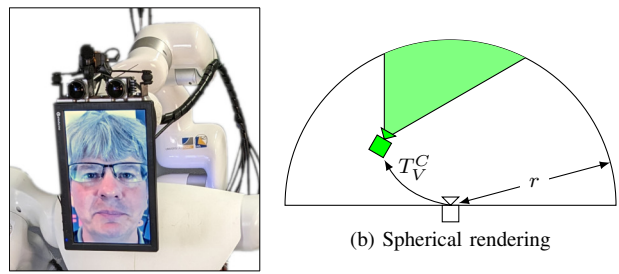}};
  \begin{scope}[shift={(4.2,0)},scale=1.1]
   \begin{scope}
    \draw (2,0) arc (0:180:2) -- cycle;
    \clip (2,0) arc (0:180:2) -- cycle;

    \begin{scope}[shift={(-0.8,0.8)}, rotate=-30]
       \draw[fill=green!50]
       (0,0) -- (60:4) arc [radius=4,start angle=60, delta angle=60] -- cycle;
    \end{scope}
    \draw[latex-latex] (15:0.1) -- node[midway,fill=white] {$r$} (15:2);
   \end{scope}
   \draw[fill=white]
       (-0.1,-0.2) rectangle (0.1,0.0)
       (0.0,0.0) -- ++(-0.1,0.1) -- ++(0.2,0.0) -- cycle;

   \begin{scope}[shift={(-0.8,0.8)}, rotate=-30]
   \draw[fill=green]
       (-0.1,-0.2) rectangle (0.1,0.0)
       (0.0,0.0) -- ++(-0.1,0.1) -- ++(0.2,0.0) -- cycle;
   \end{scope}

   \draw[-latex] (0,0) to [in=-60,out=170] node [pos=0.9,right] {$T^C_V$} (-0.8,0.55);

   \node[anchor=north] at (0,-0.2) {(b) Spherical rendering};
  \end{scope}
 \end{tikzpicture}
 \caption{Avatar robot head \& Spherical rendering.
  (a) The NimbRo avatar robot head with a 6-DoF neck and stereo cameras.
  (b) Spherical rendering example. We show only one camera $C$ of the stereo pair,
   the other is processed analogously.
   The robot camera is shown in white with its very wide FoV. The corresponding VR camera $V$,
   which renders the view displayed to the operator, is shown in green.
   The camera image is projected onto the sphere with radius $r$, and then back into the VR camera. Adapted from \citet{schwarz2021low}.}
   \label{fig:spherical}
\end{figure}

A unique property of the NimbRo Avatar system is the movable 6-DoF head, which
follows the operator's head motion with 1:1 correspondence.
This feature generates proper parallax and allows looking around objects
and easily moving to new viewpoints without repositioning the robot.
However, there are latencies in the system from network transmission and
the 6-DoF neck mechanism moving the avatar's head.
This latency is compensated by rendering new views with low latency
on the operator side under the assumption of constant distance (see \cref{fig:spherical}).
In essence, each video frame is rendered by projection onto a sphere
centered onto the camera position at time of capture, allowing the
VR camera to move freely in this sphere.
This approach hides movement latency and is unnoticeable except for
large head translations which reveal wrong distance assumptions.
Note that it is beneficial to have substantially larger field of view in the camera system ($>\!180^\circ$)
than in the HMD ($108^\circ$) in this setup, since the operator can turn their head quickly and
would see past the field of view in that case, breaking immersion.
We refer to \citet{schwarz2021low} for further details and a user study
which shows the benefit of the movable head compared to standard
pan/tilt mounting and fixed mounting.

\begin{figure}
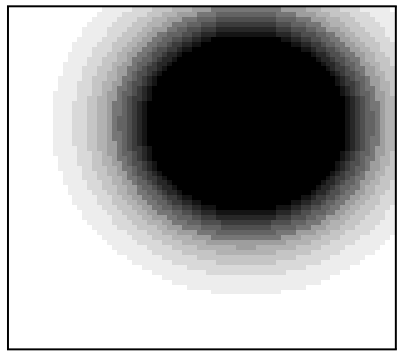

 \centering
 \includenocompress[frame,width=.6\linewidth]{images/fov/focus.png}
 \caption{Exemplary QP modulation mask for an image of size 2472$\times$2178,
 resulting in a mask size of 78$\times$69. White corresponds to
 a high QP change (yielding higher compression) while black indicates no QP change.
 }
 \label{fig:focus}
\end{figure}

\subsection{Quantization Parameter Modulation}

The Quantization Parameter (QP) of the HEVC codec directly determines the quantization
of individual blocks and thus influences the compression factor.
An increase of the QP by one corresponds to a 12\% increase in the quantization step size,
resulting in 12\% decrease of the bitrate for the same content.
NVENC allows specifying a \textit{QP delta} $\Delta q$ for each macroblock,
which in our case corresponds to 32$\times$32 pixels.

We choose to use a spatial Gaussian to produce $\Delta q$. While this
does not closely match the behavior of human visual acuity~\citep{wang2023foveated},
it allows us to easily define a wider foveal area that tolerates errors
in eye tracking as well as latencies introduced by network transmission
and camera frame rate. We thus define
\begin{equation}
 \Delta q (p) = A \left[ 1 - \exp \left( - \frac{||p - \mu||_2^2}{2 \sigma} \right) \right ],
\end{equation}
where $p$ is the normalized location in the image, $\mu$ is the projected gaze direction,
and $\sigma$ and $A$ control spread and strength of the quality reduction, respectively.
\cref{fig:focus} shows an exemplary QP mask.
We recommend $A=50$ and $\sigma=0.03$ (see \cref{sec:user}).

\subsection{Foveated Warping}

\begin{figure}
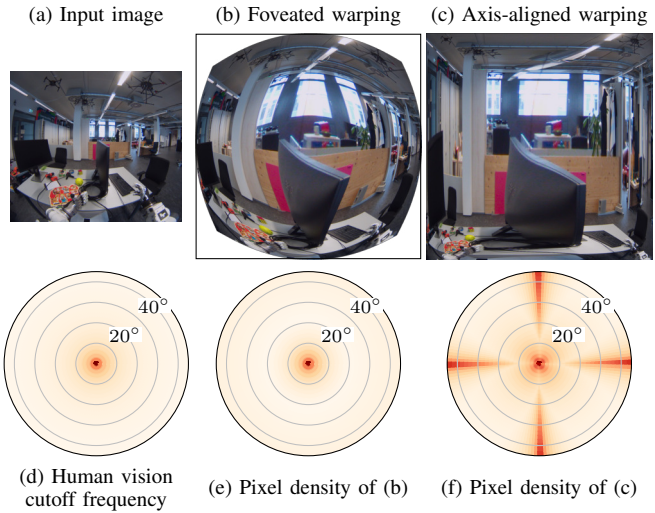

 \centering
 \begin{tikzpicture}[font=\rmfamily\footnotesize]
  \matrix[matrix of nodes, inner sep=0pt, row sep=1pt, column sep=2pt, nodes={anchor=center}] {
    (a) Input image & (b) Foveated warping & (c) Axis-aligned warping \\[2pt]
    \includegraphics[height=2cm]{images/warp/input.png} &
    |[border on top]| \includegraphics[height=3cm]{images/warp/accurate.png} &
    \includegraphics[height=3cm]{images/warp/individual.png} \\[3pt]
    \includegraphics{images/warp/density_hvs.pdf} &
    \includegraphics{images/warp/density_accurate.pdf} &
    \includegraphics{images/warp/density_individual.pdf} \\[3pt]
    \node[align=center] {(d) Human vision\\cutoff frequency}; & (e) Pixel density of (b) & (f) Pixel density of (c) \\
  };
 \end{tikzpicture}
 \caption{
   Foveated warping example. The input image (a) is warped so that pixel density, visualized in
   a polar fixation-centric coordinate system (e) follows the cutoff frequency of the human vision
   system (d). However, this wastes space in the corners.
   By restricting
   the warping to individual axes, the entire space can be utilized (c), however this introduces
   additional modes in pixel density (f).
   In the lower plots, darker color indicates higher frequency or density.
 }
 \label{fig:warping}
\end{figure}

For qualitative and quantitative comparisons, we also implement an alternative method for foveated bandwidth reduction
called \textit{Foveated Warping}~\citep{illahi2021foveated},
where images are transformed in such a way that areas close to the gaze fixation point are
represented with more pixels, while peripheral areas receive less space in the image.
Consequently, the total image resolution can be reduced, which will reduce video bandwidth automatically.

While having the advantage that the local compression effect can be modulated with higher spatial
resolution (unlimited by the macroblock size), this approach also has drawbacks:
The warping introduces additional sampling effects and requires quasi-random memory access.
Furthermore, the resulting video stream is not compatible with standard viewers and requires
de-warping---whereas the QP modulation method does not need any modification on the receiving side.

Nevertheless, we can apply foveated warping to our robotic telepresence scenario.
In contrast to existing work, which mostly focuses on VR applications~\citep{illahi2021foveated},
our input images have fisheye characteristics, which have to be considered for warping.

As in related work~\citep{illahi2021foveated}, we use the model introduced by \citet{geisler1998real}
for the cutoff frequency of the human vision system (HVS) depending on the angle $\theta$ (in degrees) to the fixation:
\begin{equation}
 f(\theta) = \frac{e_2}{\theta + e_2} \frac{1}{\alpha} \ln \left( \frac{1}{\mathrm{CT}_0} \right),
\end{equation}
where $e_2=2.3$, $\alpha=0.106$, and $\mathrm{CT}_0=\frac{1}{64}$ are the model parameters.

We then define the desired pixel density $g(\theta)=4 f(\theta)$, which ensures that the sampling theorem is satisfied (with margin for error). Finally, integrating over $\theta$ gives
\begin{equation}
 G(\Theta) = \int_0^\Theta g(\theta) d\theta,
\end{equation}
the required pixels for the angle range up to $\Theta$.

For each output pixel $(x,y)$ (centered around the fixation point) we can then compute $\theta_{(x,y)} = G^{-1}(||(x,y)||_2)$, the corresponding angle to the fixation point. Using $\theta_{(x,y)}$ the relevant 3D ray can be calculated,
which is finally projected to the source image using the camera model.

A resulting warping can be seen in \cref{fig:warping}. To fully utilize the rectangular input shape of the
video codec, the warping can be done for each axis individually. This deviates from the desired pixel density (which, however, is still a minimum bound).

\section{Experiments}

We integrate our method in the NimbRo avatar system
and report qualitative and quantitative experiments.

\subsection{Bandwidth Reduction}

 \pgfplotscreateplotcyclelist{custom mark list}{
    every mark/.append style={solid,fill=\pgfplotsmarklistfill},mark=*\\
    every mark/.append style={solid,fill=\pgfplotsmarklistfill},mark=square*\\
    every mark/.append style={solid,fill=\pgfplotsmarklistfill},mark=triangle*\\
    every mark/.append style={solid,fill=\pgfplotsmarklistfill},mark=diamond*\\
    every mark/.append style={solid,fill=\pgfplotsmarklistfill!40},mark=otimes*\\
    every mark/.append style={solid},mark=|\\
    every mark/.append style={solid,fill=\pgfplotsmarklistfill},mark=pentagon*\\
    every mark/.append style={solid},mark=text,text mark=p\\
    every mark/.append style={solid},mark=text,text mark=a\\
}

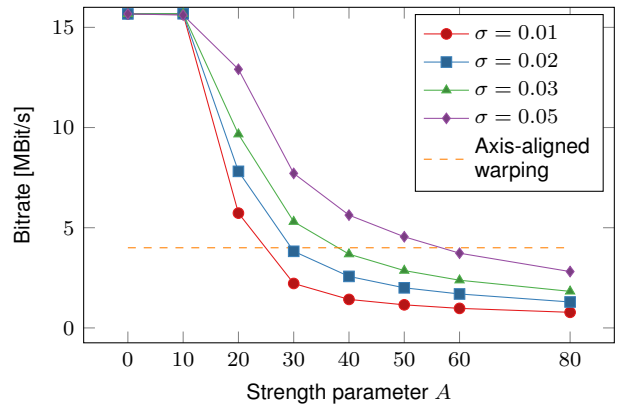
\begin{figure}
  \begin{tikzpicture}
    \begin{axis}[domain=0:80,ymax=16,ylabel={Bitrate [MBit/s]},xlabel={Strength parameter $A$},xtick=data,width=\linewidth,height=.7\linewidth,
    cycle list/Set1-8,
    cycle multiindex* list={
      custom mark list\nextlist
      Set1-8\nextlist
    },
    legend style={cells={align=left,anchor=west}},
    ]
      \addlegendentry{$\sigma=0.01$}
      \addplot+[
      ] table [x=A,y=mbits] {experiments/new0.01.txt};

      \addlegendentry{$\sigma=0.02$}
      \addplot+[
      ] table [x=A,y=mbits] {experiments/new0.02.txt};

      \addlegendentry{$\sigma=0.03$}
      \addplot+[
      ] table [x=A,y=mbits] {experiments/new0.03.txt};

      \addlegendentry{$\sigma=0.05$}
      \addplot+[
      ] table [x=A,y=mbits]
        {experiments/new0.05.txt};

      \addlegendentry{Axis-aligned\\warping}
      \addplot+[no marks,dashed] {4};
    \end{axis}
  \end{tikzpicture}
  \vspace{-1ex}
  \caption{Bandwidth reduction in foveated compression dependent on strength parameter $A$ and spatial spread $\sigma$. For comparison, a reasonable bitrate for the axis-aligned warping method is shown.}
  \label{fig:bandwidth}
\end{figure}

Since the main motivation is bandwidth reduction, we measure bandwidth and show results in \cref{fig:bandwidth}.
We can see that bandwidth decreases linearly with rising strength parameter $A$ at first, but there are diminishing returns---which is expected,
since the foveal region needs to be transmitted at original quality and thus consumes constant bandwidth.
A smaller $\sigma$ also leads to small bandwidth savings, since less area is transmitted at high fidelity.

We note that the relationship between $\Delta q$ and bitrate/bandwidth is complex, as initial QP values
depend on scene complexity and content of each macroblock.

The comparison method (Foveated Warping) reduces the resolution from 2472$\times$2178 (5.4\,MP) to 1100$\times$1100 (1.2\,MP), greatly reducing the number of pixels. However, the warped regions contain high frequencies, which again
increases the required video bitrate. We find that 4\,MBit/s gives reasonable quality using the NVENC encoder.

\subsection{Latency}

We measure round-trip latency from an eye tracking measurement on the operator side to a video frame decoded on the operator side
with the corresponding QP mask as 50\,ms, which is in the same order of magnitude as saccadic omission~\citep{campbell1978saccadic}, a neural mechanism inhibiting processing in the human visual system during rapid eye movements. Our visual system returns to its full capacity 60\,ms after a saccade ends~\citep{Loschky2007}, giving time to update the display without noticeable latency.

The latency measurement was performed with Gigabit Ethernet connection between operator station and robot. Latency will
increase with wireless connection ($<\!10$\,ms additional latency) and with connection over the Internet (up to 300\,ms), depending on network and distance.

\begin{figure}
 \centering
 \begin{tikzpicture}
  \begin{axis}[
     colorbar horizontal,
     colorbar style={
        at={(0.5,1.1)},
        anchor=south,
        xticklabel pos=upper,
     },
     axis on top, ymin=0, ymax=95, xtick=data, ytick=data, scaled ticks=false, xticklabel style={
        /pgf/number format/fixed
},
   xlabel={Spatial spread $\sigma$},
   ylabel={Strength parameter $A$},
   width=.9\linewidth,
   height=.65\linewidth,
   xmin=0.005,xmax=0.06,
   every colorbar/.append style={
     xtick={0,1,2},
     xticklabels={0: not noticeable,1: noticeable,2: inconvenient},
   },
    colormap name=greenred
   ]
   \addplot+[
    matrix plot*,point meta=explicit,mesh/cols=4,no marks,
    nodes near coords,nodes near coords align={anchor=center},
    nodes near coords style={fill=black!10},
   ] table [x=sigma,y=A,meta=rating,col sep=space] {experiments/study.csv};
  \end{axis}
 \end{tikzpicture}
 \vspace{-1ex}
 \caption{User rating experiment. For each pair of ($\sigma$,$A$) the mean user rating is visualized. A rating of zero corresponds to ``I did not notice any effect'', one means ``noticeable'', and two means ``inconvenient''.}
 \label{fig:study}
\end{figure}
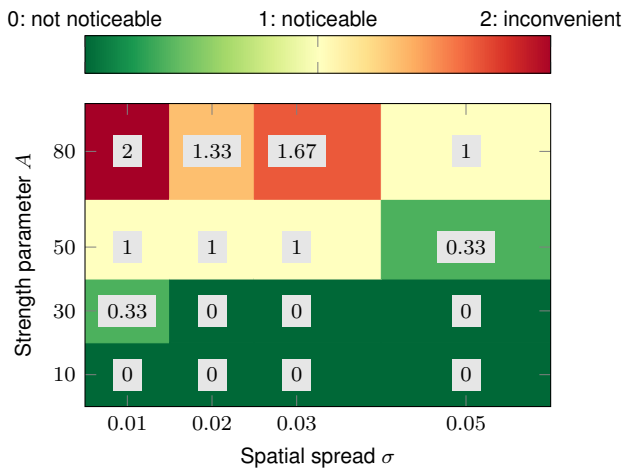

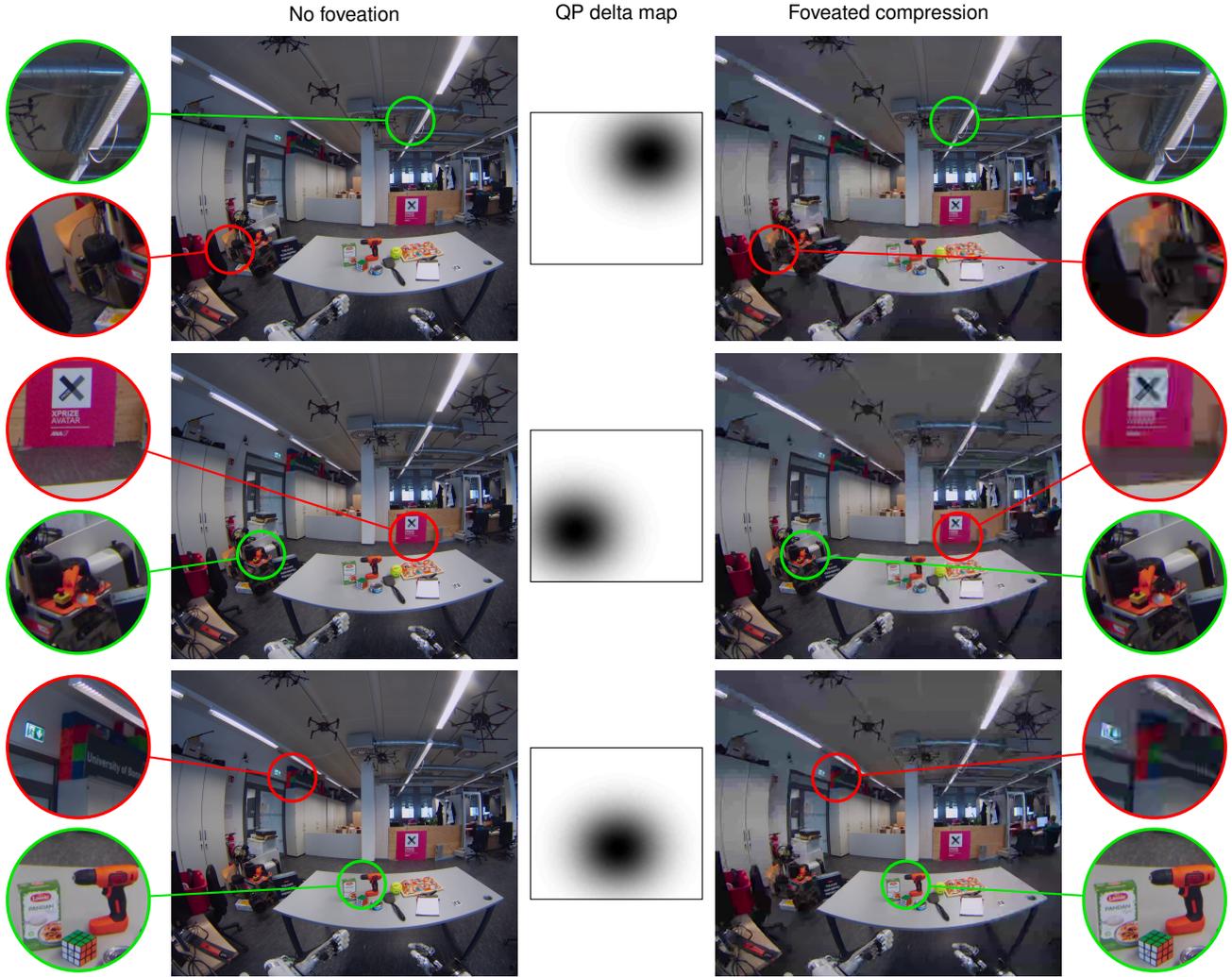
\begin{figure*}
 \centering
 \newlength\expheight\setlength\expheight{4.3cm}
 \newcommand{\spycompa}[3][]{%
   \spy[#1] on (rel cs:name=i#2,#3) in node [anchor=east,xshift=-0.3cm] at ($(i#2.north west)!0.5!(i#2.west)$);
   \spy[#1] on (rel cs:name=o#2,#3) in node [anchor=west,xshift=0.3cm]  at ($(o#2.north east)!0.5!(o#2.east)$);
 }
 \newcommand{\spycompb}[3][]{%
   \spy[#1] on (rel cs:name=i#2,#3) in node [anchor=east,xshift=-0.3cm] at ($(i#2.south west)!0.5!(i#2.west)$);
   \spy[#1] on (rel cs:name=o#2,#3) in node [anchor=west,xshift=0.3cm]  at ($(o#2.south east)!0.5!(o#2.east)$);
 }
 \begin{tikzpicture}[
    every outer matrix/.append style={inner sep=0pt},
    negative/.style={draw=red},
    positive/.style={draw=green!90!black},
    spy using outlines={circle, positive, very thick, magnification=3,
                       size=2cm, connect spies,
                       every spy in node/.append style={very thick},
                       every spy on node/.append style={very thick},
                       spy connection path={\draw[thick] (tikzspyonnode) -- (tikzspyinnode);}}
  ]
  \matrix[
    matrix of nodes,
    nodes={inner sep=0pt, anchor=center},
    column sep=5pt, row sep=5pt,
    column 2/.style={every node/.append style={border on top}},
    row 1/.style={anchor=base}
  ] {
    No foveation                                             & |[border on top prevent]| QP delta map                      & Foveated compression \\
    |(i1)| \includenocompress[height=\expheight]{experiments/table_no_focus.jpg} & |(f1)| \includenocompress[height=.5\expheight]{experiments/table50_focus_0.69_0.28.focus.png} & |(o1)|\includenocompress[height=\expheight]{experiments/table50_focus_0.69_0.28.jpg} \\
    |(i2)| \includenocompress[height=\expheight]{experiments/table_no_focus.jpg} & |(f2)| \includenocompress[height=.5\expheight]{experiments/table50_focus_0.26_0.66.focus.png} & |(o2)|\includenocompress[height=\expheight]{experiments/table50_focus_0.26_0.66.jpg} \\
    |(i3)| \includenocompress[height=\expheight]{experiments/table_no_focus.jpg} & |(f3)| \includenocompress[height=.5\expheight]{experiments/table50_focus_0.51_0.66.focus.png} & |(o3)|\includenocompress[height=\expheight]{experiments/table50_focus_0.51_0.66.jpg} \\
  };

  \spycompa{1}{x=0.69,y=0.72}
  \spycompb[negative]{1}{x=0.17,y=0.3}

  \spycompa[negative]{2}{x=0.7,y=0.4}
  \spycompb{2}{x=0.26,y=0.34}

  \spycompa[negative]{3}{x=0.35,y=0.65}
  \spycompb{3}{x=0.55,y=0.3}
 \end{tikzpicture}
 \caption{Qualitative examples of Foveated Compression.
 We show enlarged views of points close to the gaze direction (green) and points in the peripheral area (red).
 All examples use the recommended values of $A=50$, $\sigma=0.03$.
 }
 \label{fig:qualitative_table}
\end{figure*}

\begin{figure}
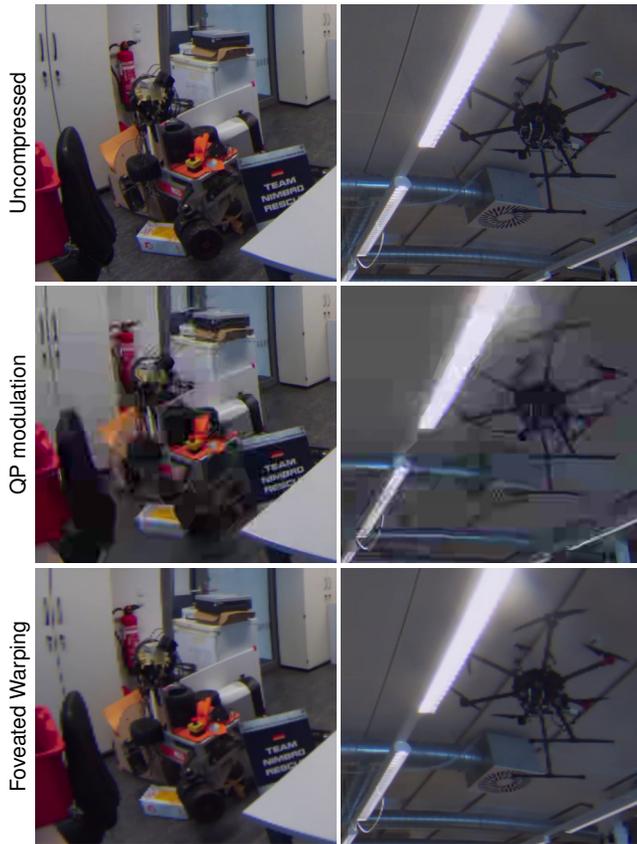

 \centering
 \begin{tikzpicture}
 \matrix[matrix of nodes, inner sep=0pt, row sep=2pt, column sep=2pt,nodes={anchor=center}] {
 |[rotate=90,anchor=base]| Uncompressed &
 \includegraphics[width=4cm]{experiments/table_no_focus.crop1.jpg} &
 \includegraphics[width=4cm]{experiments/table_no_focus.crop2.jpg} \\
 |[rotate=90,anchor=base]| QP modulation &
 \includegraphics[width=4cm]{experiments/table50_focus_0.51_0.66.crop1.jpg} &
 \includegraphics[width=4cm]{experiments/table50_focus_0.51_0.66.crop2.jpg} \\
 |[rotate=90,anchor=base]| Foveated Warping &
 \includegraphics[width=4cm]{images/dewarped_crop1.png} &
 \includegraphics[width=4cm]{images/dewarped_crop2.png} \\
 };
 \end{tikzpicture}
 \caption{Quality of QP modulation compared to Foveated Warping. Both crops are far away from the fixation point and
 demonstrate maximum quality loss.}
 \label{fig:qualitative_warp}
\end{figure}

\subsection{Qualitative Examples}

We show qualitative examples in \cref{fig:qualitative_table}.
The effect of our method can clearly be seen as content near the gaze direction is preserved and quality is drastically reduced in the periphery. Compared to Foveated Warping, blocking artifacts can appear,
although far away from the fixation point (see \cref{fig:qualitative_warp}).

\subsection{User Rating}
\label{sec:user}

To investigate which parameter settings of $A$ and $\sigma$ are acceptable for users, we performed an experiment
with three inexperienced operators who where asked to rate specific settings with the following options:
A rating of 0 means that there is no noticeable difference to the baseline (i.e. no foveation). Next, one
means there is a noticeable difference, and two says that the effect is inconvenient or breaks immersion.
As can be seen in \cref{fig:study}, larger settings of $\sigma$ allow higher settings of $A$, as expected.

Problematic scores for low $\sigma$ mostly resulted from limited eye tracking precision. In these cases,
the true operator focus point was too far away from the foveal area defined by
eye tracking and $\sigma$.
Eye tracking precision may be limited due to
slight slippage of the HMD after calibration. We recommend setting $\sigma \geq 0.03$ to avoid these issues.

\subsection{Limitations}

Increasing the strength parameter $A$ too much will result in high compression rates, which will of course lead to
artifacts at some point.
Since each macroblock is quantized separately, the macroblock grid will become visible at high QP offsets
(see \cref{fig:qualitative_warp})---though this is usually not noticeable in peripheral vision.

The NimbRo avatar system uses periodic intra refresh instead of IDR frames, i.e. instead of transmitting
full self-contained frames in periodic intervals,
the intra refresh method sweeps across the image and sends self-contained data only for a subset of the macroblocks
at a time. However, we noticed that at high QP settings this sweep becomes visible and may be noticed as movement in the peripheral vision.
We note that IDR frames are not a viable alternative, since they either lead to highly unstable transmission bitrate
or, if bitrate is kept constant, result in a noticeable flash of lower-quality content across the entire frame.
It would be conceivable to allocate more bits to the macroblocks affected by intra-refresh, but this does not seem to be possible with Nvidia NVENC.
In early experiments, patching the NVENC encoder to make the selection of macroblocks for intra-refresh more random
than the original linear sweep across the frame makes this effect less noticable.

If very high compression settings are desired, explicitly blurring peripheral regions after decompression could also reduce block effects and noticeable intra refresh.

\section{Conclusion}

We demonstrated a lightweight method for foveated compression of VR televisualization video streams. We showed
successful integration with the HEVC video codec in the NimbRo avatar system.
Qualitative and quantitative results indicate that bandwidth can be reduced to a third without noticeable artifacts that would reduce immersion.

For even higher bandwidth savings, the proposed method could be combined with Foveated Warping, which retains better
quality far away from the fixation point.
Another promising direction of future research could be the prediction of the fixation point during a saccade to mitigate latency.

\section*{Acknowledgment}

This work was supported by the Robot Industry
Core Technology Development Program (20023294,
Development of shared autonomy control framework and
AI-based application technology for enhancing tasks of
hyper realistic telepresence robots in unstructured
environment) funded by the Korean Ministry of Trade, Industry \&
Energy (MOTIE).

\printbibliography

\end{document}